\documentclass[preprint,aps,showpacs]{revtex4}
\usepackage{graphicx,epsf}
\usepackage{lscape}
\usepackage{citesort}
\usepackage{amsmath}

\begin{document}

\title{{\Large{\bf Form factors and branching ratios of the FCNC $B \to a_1 \ell^{+}\ell^{-}$ decays }}}

\author{\small
R. Khosravi \footnote {e-mail: rezakhosravi @ cc.iut.ac.ir } }

\affiliation{Department of Physics, Isfahan University of
Technology, Isfahan 84156-83111, Iran }

\begin{abstract}
We analyze the semileptonic  $B \to a_1
\ell^+\ell^-$, $\ell=\tau, \mu, e$  transitions in
the frame work of the three-point QCD sum rules in the standard
model. These rare decays governed by  flavor-changing neutral current transition of $b \to d$.
Considering the quark condensate contributions, the relevant form factors as well as
the branching fractions of these transitions are calculated.
\end{abstract}


\maketitle

\section{Introduction}
The decays governed by flavor-changing neutral current (FCNC)
transitions are very sensitive to the gauge
structure of the standard model (SM) which  provide an excellent way to test such a model.
These decays, prohibited at the tree-level, take place at loop level by electroweak
penguin and weak box diagrams.
The FCNC transitions can be suppressed due to their proportionality
to the small  Cabibbo-Kobayashi-Maskawa matrix elements (for instance see \cite{DP}).
Among these, the FCNC semileptonic decays of the $B$ meson occupy a special place
in both experimental measurements and theoretical
studies for the precision test of the SM due to more simplicity.

So far, the form factors of the semileptonic decay $B \to a_1 \ell \nu $  have been studied via the different approaches such as the covariant
light front quark model (LFQM) \cite{CLF04},
the constituent quark-meson model (CQM) \cite{QM99},  the light cone QCD sum rules (LCSR) \cite{LCSR07},
 and the QCD sum rules (SR) \cite{Aliev99}. However, the obtained results of these methods are different from each other.

In this work, we calculate
the transition form factors of the FCNC semileptonic decays  $B \to a_1(1260) \ell^{+} \ell^{-}/ \nu \bar{\nu}$
in the framework of the three-point QCD sum rules method (3PSR).
Considering the transition  form factors for such decays in the framework of
different theoretical methods has two-fold importance:

1) A number of the physical observables such as branching ratio, the
forward-backward asymmetry and lepton polarization asymmetry,  which have  important roles in testing the SM
and  searching for new physics beyond the SM, could be investigated.

2) These form factors can be also  used  to determine the
factorization  of amplitudes in the non-leptonic  two-body
decays.

On the other hand, any experimental measurements of the present quantities
and a comparison with the theoretical predictions can give
valuable information about the FCNC transitions and strong interactions in $B \to a_1 \ell^+ \ell^-/\nu
\bar{\nu}$ decays.

The plan of the present paper is as follows: In Sec. II, we
describe  the sum rules method to calculate the  form factors of the FCNC $B \to a_1 $  transition.
Section III is devoted to the numerical analysis of the form factors and  branching ratio values of
the semileptonic  $B \to a_1$ decays, with and without the long-distance (LD) effects.

\section{Form factors of the FCNC $B\to a_1$ transition in 3PSR}
In the SM, the rare semileptonic decays which occur via
$b \to d~  \ell^+ \ell^-$ transition is described by the effective Hamiltonian as \cite{Khodjamirian}:
\begin{equation}
H_{\rm eff} = - \frac{G_F}{\sqrt{2}} V_{tb}V_{td}^{*}
              \sum_{i=1}^{10} C_i(\mu)  O_i(\mu) \; .
\end{equation}
where $V_{tb}$ and $V_{td}$ are the elements
of the CKM matrix, and $C_i(\mu)$ are the Wilson coefficients. It should be noted that the CKM-suppressed contributions proportional to $V_{ub}V^*_{ud}$ is neglected, also
the approximation $|V_{tb}V^*_{td}|\simeq|V_{cb}V^*_{cd}|$ is adopted \cite{Charles}.
The standard set of the local operators for $b \to d \ell^+ \ell^-$ transition is written as \cite{FGIKL}:
\begin{eqnarray}
\begin{array}{ll}
O_1     =  (\bar{d}_{i}  c_{j})_{V-A} ,
           (\bar{c}_{j}  b_{i})_{V-A}  ,               &
O_2     =  (\bar{d} c)_{V-A}  (\bar{c} b)_{V-A}   ,              \\
O_3     =  (\bar{d} b)_{V-A}\sum_q(\bar{q}q)_{V-A}  ,            &
O_4     =  (\bar{d}_{i}  b_{j})_{V-A} \sum_q (\bar{q}_{j}
          q_{i})_{V-A} ,                                    \\
O_5     =  (\bar{d} b)_{V-A}\sum_q(\bar{q}q)_{V+A} ,            &
O_6     =  (\bar{d}_{i}  b_{j })_{V-A}
   \sum_q  (\bar{q}_{j}  q_{i})_{V+A}  ,               \\
O_7     =  \frac{e}{8\pi^2} m_b (\bar{d} \sigma^{\mu\nu}
          (1+\gamma_5) b) F_{\mu\nu}  ,                     &
O_8    =  \frac{g}{8\pi^2} m_b (\bar{d}_i \sigma^{\mu\nu}
   (1+\gamma_5) { T}_{ij} b_j) { G}_{\mu\nu}  ,          \\
O_9     = \frac{e}{8\pi^2} (\bar{d} b)_{V-A}  (\bar{l}l)_V ,      &
O_{10}  = \frac{e}{8\pi^2} (\bar{d} b)_{V-A}  (\bar{l}l)_A
\end{array}
\end{eqnarray}
where ${ G}_{\mu\nu}$ and $F_{\mu\nu}$ are the gluon and photon
field strengths, respectively; ${T}_{ij}$ are the generators of
the $SU(3)$ color group; $i$ and $j$ denote color indices.
Labels $(V\pm A)$ stand for
$\gamma^\mu(1\pm\gamma^5)$.  $O_{1,2}$ are current-current operators,
$O_{3-6}$ are QCD penguin operators,  $O_{7,8}$ are magnetic
penguin operators, and $O_{9,10}$ are semileptonic electroweak penguin
operators.

\begin{figure}[th]
\includegraphics[width=10cm,height=3cm]{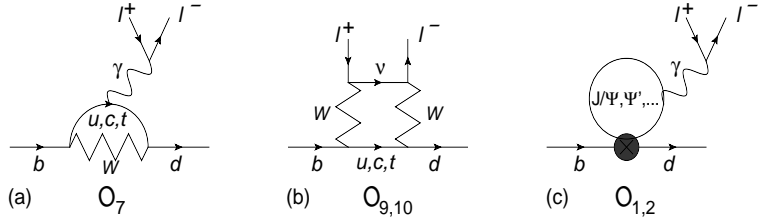}
\caption{(a) and (b) $O_7$ and $O_{9,10}$ short distance contributions.   (c) $O_{1,2}$ long distance charm-loop contribution.} \label{F21}
\end{figure}

The most relevant contributions to $B \rightarrow
a_1 \ell^+ \ell^-$ transitions are given by the
$O_7$ and  $O_{9,10}$, short distance (SD) contributions, as well as the tree-level four quark operators
$O_{1,2}$ which have sizeable Wilson coefficients.
The current-current operators $O_{1,2}$ involves an intermediate charm-loop, LD contributions, coupled to the lepton pair via the virtual photon (see Fig. \ref{F21}).
This contribution has got the same form factor dependence as $C_9$ and can therefore be absorbed into an effective Wilson coefficient
$C^{\rm eff}_9$ \cite{LYON}.

Therefore,  the effective Hamiltonian for $B \rightarrow
a_1 \ell^+ \ell^-$  decays which occur via $b \rightarrow
d \ell^+ \ell^-$ loop  transition   can be written as:
\begin{eqnarray}
H_{\rm eff}&=& \frac{G_{F}\alpha}{2\sqrt{2}\pi} V_{tb}V_{td}^{*}\Bigg[
C_9^{\rm eff}  \overline {d} \gamma_\mu (1-\gamma_5) b~  \overline {l
}\gamma_\mu l + C_{10} \overline {d} \gamma_\mu (1-\gamma_5) b~
\overline {l } \gamma_\mu \gamma_5 \l \nonumber\\&-& 2 C_7^{\rm eff}\frac{m_b}{q^2}
\overline {d} ~i\sigma_{\mu\nu}
q^\nu (1+\gamma_5) b~  \overline {l}  \gamma_\mu l \Bigg],
\end{eqnarray}
where $C_7^{\rm eff}= C_7-C_5/3-C_6$. The effective Wilson coefficients $C^{\rm eff}_9(q^2)$, are given as
\begin{eqnarray}\label{eq33}
C^{\rm eff}_9(q^2) = C_9 + Y(q^2)\,.
\end{eqnarray}
The function $Y(q^2)$ contains the LD
contributions coming from the real $c\bar c$ intermediate states called charmonium resonances. Two resonances, $J/\psi$ and  $\psi'$, are narrow and the
last four resonances, $\psi(3370)$, $\psi(4040)$, $\psi(4160)$ and $\psi(4415)$, are above the $D\bar D$-threshold and as a consequence the width is much larger. The
explicit expressions of the $Y(q^2)$ can be
found in \cite{LYON} (see also \cite{BM, FGIKL}).

To calculate the form factors of the FCNC $B \to a_1$ transition,
within 3PSR method, we start with the following correlation
functions constructed from the transition currents $J^{V-A}_{\mu}=\bar
d \gamma_\mu (1-\gamma_5) b$ and $J^{T}_{\mu}=\bar d\, i
\sigma_{\mu\eta}q^\eta (1+\gamma_5) b$ as follows:
\begin{eqnarray}\label{eq21}
{\Pi}_{\mu\nu}^{V-A~(T)}(p^2,p'^2,q^2)&=&\int
d^{4}xd^4ye^{-ipx}e^{ip'y}\langle0\mid \mathcal{T}[J _{\nu}^{a_1}(y)
J_{\mu}^{V-A~(T)}(0) {J^{B}}^{\dag}(x)]\mid  0\rangle\,,
\end{eqnarray}
where  $J^{B}= \bar u  \gamma_5 b$, and $J^{a_1}_{\nu}=\bar u \gamma_\nu \gamma_5
d$ are the interpolating
currents of the initial and final meson states, respectively.
In the QCD sum rules approach, we can obtain the
correlation functions of Eq. (\ref{eq21}) in two languages: the
hadron language, which is the physical or phenomenological
side, and the quark-gluon language called the QCD
or theoretical side. Equating two sides and applying the
double Borel transformations with respect to the momentum
of the initial and final states to suppress the contribution of
the higher states and continuum, we get sum rule expressions
for our form factors. To drive the phenomenological
part, two complete sets of intermediate states with the same
quantum numbers as the currents $J^{a_1}_{\nu}$ and $J^B$ are inserted in Eq. (\ref{eq21}). As a result of this procedure\,,
\begin{eqnarray}\label{eq22}
\Pi^{V-A~(T)}_{\mu\nu} (p,p^\prime) &=& \frac{1}{p^2-m_B^2} \,
\frac{1}{p^{\prime 2}-m_{a_1}^2} \langle 0 | J^{a_1}_{\nu} | a_1 \rangle \langle a_1 | J^{V-A~(T)}_{\mu}| B \rangle \langle B | {J^{B}}^{\dag} | 0
\rangle + \mbox{higher states\,,}\,\nonumber\\
\end{eqnarray}
where $p$ and $p^{\prime}$ are the momentum of the initial and final
meson states, respectively. To get the transition matrix elements of the $B \to a_1$  with various quark models, we
parameterize them in terms of the relevant form factors as
\begin{eqnarray}\label{eq23}
\langle a_1(p',\epsilon)\mid J^{V-A}_{\mu}\mid
B(p)\rangle&=&\frac{1}{m_{B}+m_{a_1}}\,\left[2\,A(q^2)\,i\varepsilon_{\mu\lambda\alpha\beta}
\epsilon^{\ast\lambda}p^\alpha p'^\beta+\,{V}_{1}(q^2)(P.q)
\epsilon_{\mu}^{\ast}\right.\nonumber\\&+&\left. {V}_{2}(q^2)\,(\epsilon^{*}. p)P_{\mu}
+\,{V}_{0}(q^2)\,(\epsilon^{*}. p) q_{\mu}\right]\,,\nonumber\\
\langle a_1(p',\epsilon)\mid J^{T}_{\mu}\mid
B(p)\rangle&=&2~{T}_{1}(q^2)~i\varepsilon_{\mu\lambda\alpha\beta}
\epsilon^{\ast\lambda}p^\alpha p'^\beta+
{T}_{2}(q^2) (m_{B}^2-m_{a_1}^2)\left[\epsilon_{\mu}^{\ast}
-\left. \frac{1}{q^2} (\epsilon^{*}. q)q_{\mu}\right]
\right.\nonumber\\&+&{T}_{3}(q^2)~(\epsilon^{*}. p)\left[P_{\mu}-\frac{1}{q^2}(P. p)q_{\mu}\right]\,,
\end{eqnarray}
where $P=p+p'$ and $q=p-p'$. Also $m_{a_1}$ and $\epsilon$ are the mass and the
four-polarization vector of the $a_1$ meson.  The vacuum-to-meson
transition matrix elements are  defined in standard way, namely
\begin{eqnarray}\label{eq24}
\langle 0 | J^{B} | B \rangle =-i f_B
\frac{m_B^2}{m_b}\,,\quad\quad \langle 0 | J_{\nu}^{a_1} | a_1 \rangle = f_{a_1}m_{a_1}
\epsilon_\nu~.
\end{eqnarray}
Using Eq. (\ref{eq23}), and Eq. (\ref{eq24}) in Eq.
(\ref{eq22}), and performing summation over the polarization of
the $a_1$ meson, we obtain
\begin{eqnarray}\label{eq25}
\Pi_{\mu\nu}^{V-A}&=&\frac{f_{B}m_{B}^2}{m_{b}}\frac{f_{a_1}m_{a_1}}
{(p^2-m_{B}^2)(p'^2-m_{a_1}^2)} \times \left[ \frac{2 {A}}{m_{B}+m_{a_1}}(q^2)\,\varepsilon_{\mu\nu\alpha\beta}p^{\alpha}p'^{\beta}
-i{V}_{1}(q^2)\,(m_{B}-m_{a_1})\,g_{\mu\nu}\right.\nonumber\\&-i&\left.\frac{{V}_{2}(q^2)}{m_{B}+m_{a_1}}\, P_{\mu}p_{\nu} -i\frac{{V}_{0}(q^2)}{m_{B}+m_{a_1}}\,q_{\mu}p_{\nu}\right]
+ \mbox{excited states}\,,\nonumber\\
\Pi_{\mu\nu}^{T}&=&\frac{f_{B}m_{B}^2}{m_{b}}\frac{f_{a_1}m_{a_1}}
{(p^2-m_{B}^2)(p'^2-m_{a_1}^2)} \times \left[2\,{T
}_{1}(q^2)\,\varepsilon_{\mu\nu\alpha\beta}p^{\alpha}p'^{\beta}
-i {T}_{2}(q^2)\,(m_B^2-m_{a_1}^{2})\,g_{\mu\nu}\right.\nonumber\\&-i&\left. {T}_{3}(q^2)\,P_{\mu}p_{\nu}
\right]+ \mbox{excited states}\,.
\end{eqnarray}
To calculate the form factors $A$, $V_i (i=0, 1, 2)$, and ${T}_{j} (j= 1, 2, 3)$, we will
choose the structures
$\varepsilon_{\mu\nu\alpha\beta}p^{\alpha}p'^{\beta}$,
$g_{\mu\nu}$, $P_{\mu}p_{\nu}$, $q_{\mu}p_{\nu}$, from
$\Pi_{\mu\nu}^{V-A}$ and
$\varepsilon_{\mu\nu\alpha\beta}p^{\alpha}p'^{\beta}$, $g_{\mu\nu}$,
and $P_{\mu}p_{\nu}$ from $\Pi_{\mu\nu}^{T}$, respectively.
For simplicity, the correlations are written as
\begin{eqnarray}\label{eq26}
\Pi^{V-A}_{\mu\nu}(p^2,p'^2,q^2) &=& \Pi^{V-A}_{A}\,
\varepsilon_{\mu\nu\alpha\beta} p^\alpha p^{\prime \beta}  -i  \Pi^{V-A}_{1}\, g_{\mu\nu} -i \Pi^{V-A}_{2}\,  P_\mu p_\nu
-i\Pi^{V-A}_{0}\,  q_\mu p_\nu +
\cdots\,,\nonumber\\
\Pi^{T}_{\mu\nu}(p^2,p'^2,q^2) &=& \Pi^{T}_{1}\, \varepsilon_{\mu\nu\alpha\beta} p^\alpha p^{\prime
\beta} -i \Pi^{T}_{2}\, g_{\mu\nu}   -i \Pi^{T}_{3}\, P_\mu p_\nu
+ \cdots\,.
\end{eqnarray}

Now, we consider the theoretical part of the sum rules.
For this aim, each $\Pi_{k}^{V-A~(T)}$ function is defined in
terms of the perturbative and nonperturbative parts as
\begin{eqnarray} \label{eq27}
\Pi^{V-A~(T)}(p^2,p'^2,q^2) = \Pi^{V-A~(T)}_{\rm per}(p^2,p'^2,q^2) +\Pi^{V-A~(
T)}_{\rm nonper}(p^2,p'^2,q^2)\,.
\end{eqnarray}
For the perturbative part, the bare-loop diagrams
are considered. With the help of the double dispersion
representation, the bare-loop contribution is written as
\begin{eqnarray*}\label{eq28}
{\Pi}_{\rm per}^{V-A~( T)}=-\frac{1}{(2\pi)^2}\int ds'\int ds\frac{{\rho}^{V-A~(
T)}(s,s',q^2)}{(s-p^2)(s'-p'^2)}+\textrm{ subtraction terms}\,,
\end{eqnarray*}
where $\rho$ is spectral
density. The spectral density is obtained from the usual Feynman
integral for the bare-loop by replacing $\frac{1}{p^2-m^2} \to
-2\pi i \delta(p^2-m^2)$.  After standard calculations for the
spectral densities ${\rho}^{V-A~(T)}_{k}$, where $k$ is related to each structure in Eq. (\ref{eq26}),
we have
\begin{eqnarray}\label{eq29}
{\rho}^{V-A}_{A}&=&3\,{ s'}\,\Lambda^{-3}\,\left( u-2\,\Delta \right) { m_b}\,,\nonumber\\
{\rho}^{V-A}_{0}&=&-\frac{3}{2}\,{ s'}\,\Lambda^{-5}\,\left(12\,u\Delta\,{ s'}- 4\,s{{ s'}}^{2}-2\,{u}^{2}{
s'}\,-12\,{ s'}\,{\Delta}^{2}\,{-}\,2\,s u{ s'}\,{-}\,6\,u{\Delta}^{2}\,{-}\,
{u}^{3}\,{+}\,6\,{u}^{2}\Delta\, \right) { m_b}\,,\nonumber\\
{\rho}^{V-A}_{1}&=&-\frac{3}{2}\,{ s'}\,\Lambda^{-3}\,\left( 2\,s{ s'}-2\,{\Delta}^{2}
+2\,\Delta\,u-{u}^{2} \right)\,{ m_b}\,,\nonumber\\
{\rho}^{V-A}_{2}&=&-\frac{3}{2}\,{ s'}\,\Lambda^{-5}\,\left(12\,u\Delta\,{ s'}- 4\,s{{ s'}}^{2}-2\,{u}^{2}{
s'}\,-12\,{ s'}\,{\Delta}^{2}\,{+}\,2\,s u{ s'}\,{+}\,6\,u{\Delta}^{2}\,{+}\,
{u}^{3}\,{-}\,6\,{u}^{2}\Delta\, \right) { m_b}\,,\nonumber\\
{\rho}^{T}_{1}&=&-3\,{ s'}\,\Lambda^{-3}\,\left( u-2\,\Delta \right) { m_b^2}\,,\nonumber\\
{\rho}^{T}_{2}&=&\frac{3}{2}\,{ s'}\,\Lambda^{-3}\,\left( 2\,{s}^{2}{ s'}-2\,s{\Delta}^{2}+2\,
s\Delta\,u-s{u}^{2}-4\,s{ s'}\,\Delta+su{ s'}+u{\Delta}^{2} \right)\,,\nonumber \\
{\rho}^{T}_{3}&=&\frac{3}{2}\,{ s'}\,\Lambda^{-5}\, \left( 4\,{s}^{2}{{\it s'}}^{2}+2\,u{s}^{2}{\it s'}+6\,s
u{{\it s'}}^{2}-8\,s{{\it s'}}^{2}\Delta+8\,\Delta\,us{\it s'}-4\,s{
\it s'}\,{\Delta}^{2}-7\,s{u}^{2}{\it s'}+s{u}^{3}\right.\nonumber\\&-&\left. 6\,s{u}^{2}\Delta+6
\,s u{\Delta}^{2}+6\,u{\Delta}^{2}{\it s'}-4\,{u}^{2}\Delta\,{\it s'}+ 4
\,\Delta\,{u}^{3}-5\,{u}^{2}{\Delta}^{2} \right)\,,
\end{eqnarray}
where $u=s+s'-q^2$, $\Lambda=\sqrt{u^2-4 s s'}$, and $\Delta=s-m_b^2$.

Now, the nonperturbative part contributions to the correlation
functions are discussed (Eq. (\ref{eq27})). In QCD, the three point
correlation function can be evaluated by the operator product
expansion (OPE) in the deep Euclidean region. Up
to dimension 6, the operators are determined by the contribution of
the bare-loop, and power corrections coming from dimension-3 $\langle
\bar \psi \psi \rangle$, dimension-4 $\langle G^2 \rangle$,  dimension-5 $m_0^2
\langle \bar \psi \psi \rangle$, and dimension-6 $\langle \bar \psi \psi \rangle^2$
operators \cite{Aliev99}.
The bare-loop diagrams, perturbative part of the correlation functions, are discussed before.
For the nonperturbative part contributions, our calculations show that the contributions coming from
$\left\langle G^2 \right\rangle$ and $\langle \bar \psi \psi \rangle^2$ are very small in
comparison with the contributions of dimension-$3$
and $5$ that, their contributions can be easily ignored. We introduce the nonperturbative part contributions  as
\begin{equation}\label{eq216}
{\Pi}^{V-A~(T)}_{\rm nonper}=\langle u
\bar{u}\rangle {C}^{V-A~(T)}\,,
\end{equation}
where $\langle u\bar{u}\rangle=-(0.240\pm 0.010)^3~  \mbox{GeV}^3$ \cite{PK}. After some straightforward calculations, the explicit expressions for ${C}_{k}^{V-A~(T)}$, are given as
\begin{eqnarray}\label{eq210}
{C}^{V-A}_A&=& \frac{1} {r r'}- m_0^2  \Bigg[ \frac{1}{3 r^2 r'} +
\frac{m_b^2-q^2}{3 r^2 r'^2}+
\frac{m_b^2}{2r^3 r'}  \Bigg]\,,\nonumber\\
{C}^{V-A}_0&=&\frac{1} {r r' }-m_0^2 \Bigg[\frac{1}{ r^2 r'} +
\frac{m_b^2-q^2}{3 r^2 r'^2}+
\frac{m_b^2}{2 r^3 r'} \Bigg]\,,\nonumber\\
{C}^{V-A}_1&=& \frac{(m_b^2-q^2)} {2 r r'}- m_0^2  \Bigg[ -
\frac{1}{6 r r'} + \frac{m_b^2-q^2}{6r r'^2} + \frac{3 m_b^2-4
q^2}{12 r^2 r'}+ \frac{(m_b^2-q^2)^2}{6r^2r'^2} + \frac{m_b^4 -
m_b^2 q^2}{4 r^3 r'} \Bigg]\,,\nonumber\\
{C}^{V-A}_2&=&-\frac{1} {r r' }-m_0^2  \Bigg[ \frac{1}{3 r^2 r'}
- \frac{m_b^2-q^2}{3 r^2 r'^2}- \frac{m_b^2}{2 r^3 r'}
\Bigg]\,,\nonumber\\
{C}_1^{T}&=& -\frac{m_b} {r r' }- m_0^2 \Bigg[ -\frac{m_b}{2 r^2
r'} - \frac{m_b(m_b^2-q^2)}{3 r^2 r'^2}-\frac{m_b^3}{2r^3 r'}
\Bigg]\,,\nonumber\\
{C}_2^{T} &=& \frac{(-m_b^3+m_bq^2)} {2 r r'}- m_0^2 \Bigg[ -
\frac{m_b}{4 r r'} - \frac{m_b(m_b^2-q^2)}{6r r'^2} - \frac{m_b(4
m_b^2-5 q^2)}{12 r^2 r'}- \frac{m_b(m_b^2-q^2)^2}{6r^2
r'^2}\nonumber\\&-& \frac{m_b^5 -
m_b^3 q^2}{4 r^3 r'} \Bigg]\,,\nonumber\\
{C}_3^{T}&=&\frac{m_b} {2 r r' }-m_0^2 \Bigg[\frac{2m_b}{3 r^2 r'}
+ \frac{m_b(m_b^2-q^2)}{8 r^2 r'^2}+
\frac{m_b^3}{4 r^3 r'} \Bigg]\,,
\end{eqnarray}
where $r=p^2-m_b^2$,  $r'=p'^2$, and  $m^2_0=(0.8\pm 0.2) \mbox{GeV}^2$ \cite{PK}.

The next step is to apply the Borel transformations as
\begin{eqnarray}\label{eq214}
{B}_{p^2}(M^2)(\frac{1}{p^2-m^2})^n=\frac{(-1)^n}{\Gamma(n)}
\frac{e^{-m^2/M^2}}{(M^2)^n}\,,
\end{eqnarray}
with respect to the $p^2 (p^2\to M_1^2)$ and $p'^2 (p'^2\to M_2^2)$ on the
phenomenological as well as the perturbative and nonperturbative
parts of the correlation functions and equate these two
representations of the correlations. The following sum rules for the
form factors are derived
\begin{eqnarray}\label{eq211}
A'(V'_i)(q^2)&=& -\frac{m_b }{ f_{B} m_{B}^2 f_{a_1} m
_{a_{1}}} e^{m_{B}^2/M_1^2} e^{m_{a_1}^2/M_2^2} \times
\Bigg\{-\frac{1}{4 \pi^2} \int_{0}^{s_0^\prime} ds^\prime
\int_{s_L}^{s_{0}}ds {\rho}_{A(i)}^{V-A} e^{-s/M_1^2}
e^{-s^\prime/M_2^2}  \nonumber \\ &+&
\langle u \bar{u}\rangle \times {B}_{p^2}(M_1^2){B}_{p'^2}(M_2^2) C^{V-A} _{A(i)}
\Bigg\}\,,\nonumber\\
T'_j(q^2)&=& -\frac{m_b }{ f_{B} m_{B}^2 f_{a_1} m
_{a_{1}}} e^{m_{B}^2/M_1^2} e^{m_{a_1}^2/M_2^2} \times
\Bigg\{-\frac{1}{4 \pi^2} \int_{0}^{s_0^\prime} ds^\prime
\int_{s_L}^{s_{0}}ds {\rho}_{j}^{T} e^{-s/M_1^2}
e^{-s^\prime/M_2^2}  \nonumber \\ &+&
\langle u \bar{u}\rangle \times {B}_{p^2}(M_1^2){B}_{p'^2}(M_2^2) C^{T} _{j}
\Bigg\}\,,
\end{eqnarray}
where
\begin{eqnarray*}\label{eq212}
A'(q^2)&=&\frac{2 A(q^2)}{m_{B}+m_{a_1}}\,,~\quad\quad\quad\quad\quad\quad
{V'}_{0}(q^2)=\frac{{V}_{0}(q^2)}{m_{B}+m_{a_1}}\,,
\nonumber\\
{V'}_{1}(q^2)&=&{V}_{1}(q^2)(m_{B}-m_{a_1})\,,~\quad\quad\quad
{V'}_{2}(q^2)=\frac{{V}_{2}(q^2)}{m_{B}+m_{a_1}}\,,\nonumber\\
{T'}_{1}(q^2)&=&2{T}_{1}(q^2)\,,\quad\quad\quad\quad\quad\quad\quad\quad {T'}_{2}(q^2)={T}_{2}(q^2)(m_{B}^2-m_{a_1}^2)\,,\nonumber\\
{T'}_{3}(q^2)&=&{T}_{3}(q^2)\,.
\end{eqnarray*}
$s_0$ and $s^\prime_0$ are the continuum thresholds in the $B$
and $a_1$ meson channels, respectively. $s_{L}$,
the lower limit of the integration
over $s$, is:
$m_b^2 +\frac{m_b^2}{m_b^2-q^2} s'$.

\section{Numerical analysis}
In this section, we present our numerical analysis of the form factors $%
A, V_i$, and $T_j$. We choose the values of  the  quark,
lepton, and meson masses and also the leptonic decay constants as:
$m_b=4.8~\mbox{GeV}$ \cite{KCB}, $m_{\mu}=0.105~\mbox{GeV}$,
$m_{\tau}=1.776~\mbox{GeV}$, $m_{a_1}=1.260~\mbox{GeV}$,
$m_{B}=5.280~\mbox{GeV}$ \cite{PDG}, $f_{a_1}=(238\pm 10)~\mbox{
MeV}$ \cite{KCYANG}. For the value of the $f_{B}$, we shall use $f_B
=140~\mbox{MeV}$. This value of $f_{B}$ corresponds to the case
where ${\cal O}(\alpha_s)$ corrections are not taken into account
(see \cite{Eletsky,Belyaev}).

The sum rules for the form factors contain also four auxiliary
parameters: Borel mass squares $M_{1}^2$ and $M_{2}^2$ and continuum
thresholds $s_{0}$ and $s'_{0}$. These are not physical quantities,
so the form factors as physical quantities should be independent
of them. The continuum thresholds of $B$ and $a_{1}$ mesons, $s_0$ and $s_0^\prime$ respectively,
are not completely arbitrary; these are in correlation with the energy of the first
exited state with the same quantum numbers as the considered
interpolating currents. The values of the continuum thresholds calculated from the
two--point QCD sum rules are taken to be $s_0=(35\pm 2)$ GeV$^2$ \cite{MASHAIVA} and
$s'_0=(2.55\pm 0.15)$ GeV$^2$ \cite{KCYANG}. We search for the intervals of the Borel mass parameters so that our
results are almost insensitive to their variations. One more
condition for the intervals of these parameters is the fact that the
aforementioned intervals must suppress the higher states, continuum
and contributions of the highest-order operators. In other words,
the sum rules for the form factors must converge (for more details, see \cite{Colangelo}).  As a result, we
get $8~ \mbox{GeV}^2 \le M_1^2 \le
15~ \mbox{GeV}^2$ and $2.5~ \mbox{GeV}^2 \le M_2^2 \le 4~ \mbox{GeV}^2$.

Equation (\ref{eq211}) shows the $q^2$ dependence of the form
factors in the region  where the sum rule is valid. To extend these
results to the full region, we look for parametrization of the form
factors in such a way that in the validity region of the 3PSR, this
parametrization coincides with the sum rules prediction. We use two
following sufficient parametrizations of the form factors with
respect to $q^2$ as:
\begin{equation}  \label{eq51}
F^{(1)}(q^2)=\frac{1}{1- (\frac{q^2}{m_B^2})}\sum_{r=0}^{2} b_r \left[z^r + (-1)^{r}\, \frac{r}{3}\, z^4 \right]\,.
\end{equation}
where
$z=\frac{\sqrt{t_{+}-q^2}-\sqrt{t_{+}-t_{0}}}{\sqrt{t_{+}-q^2}+\sqrt{t_{+}-t_{0}}}$,
$t_{+}=(m_{B}+m_{a_1})^2$ and
$t_{0}=(m_{B}+m_{a_1})(\sqrt{m_B}-\sqrt{m_{a_1}})^2$ \cite{BCL}, and
also
\begin{equation}  \label{eq52}
F^{(2)}(q^2)=\frac{f(0)}{1- \alpha (\frac{q^2}{m_B^2})+ \beta {(\frac{q^2}{m_B^2})}^2}\,.
\end{equation}

We evaluated the values of the parameters $b_r\, (r=1,...,3)$ of the first and $%
f(0)$, $\alpha$,  $\beta$ of the second fit function for each transition form factor of
the $B \to a_1$ decay, taking $M_{1}^2=10~\mbox{GeV}^2$ and $
M_{2}^2=3~\mbox{GeV}^2$. Tables \ref{T31} and \ref{T32} show the values of the $b_r$
and $f(0)$, $\alpha$, $\beta$ for the form factors.
\begin{table}[th]
\caption{The values of the $b_{r}$ related to $F^{(1)}(q^2)$. }\label{T31}
\begin{ruledtabular}
\begin{tabular}{cccccccccc}
\mbox{Parameter}&${A}^{(1)}$&${V}^{(1)}_{0}$&${V}^{(1)}_{1}$&${V}^{(1)}_{2}$&${T}^{(1)}_{1}$&${T}^{(1)}_{2}$&${T}^{(1)}_{3}$\\
\hline
$b_0$&$0.44$&$0.35$&$0.28$&$-0.30$&$-0.33$&$-0.21$&$0.33$\\
$b_1$&$0.80$&$1.77$&$2.80$&$-1.79$&$-0.60$&$-2.14$&$1.42$\\
$b_2$&$3.89$&$0.09$&$15.52$&$0.94$&$-2.90$&$-11.34$&$-0.04$
\end{tabular}
\end{ruledtabular}
\end{table}

\begin{table}[th]
\caption{The values of the $f(0)$, $\alpha$ and $\beta$ connected  to $F^{(2)}(q^2)$. }\label{T32}
\begin{ruledtabular}
\begin{tabular}{cccccccccc}
\mbox{Parameter}&${A}^{(2)}$&${V}^{(2)}_{0}$&${V}^{(2)}_{1}$&${V}^{(2)}_{2}$&${T}^{(2)}_{1}$&${T}^{(2)}_{2}$&${T}^{(2)}_{3}$\\
\hline
$f(0)$&$0.51$&$0.46$&$0.52$&$-0.41$&$-0.37$&$-0.37$&$0.41$&\\
$\alpha$&$0.58$&$0.37$&$-0.52$&$0.34$&$0.58$&$-0.50$&$0.44$&\\
$\beta$&$-0.39$&$-0.04$&$0.38$&$0.14$&$-0.40$&$0.48$&$-0.10$&
\end{tabular}
\end{ruledtabular}
\end{table}

So far, several authors have calculated the form factors of the
$B\to a_1 \ell \nu$ decay via the different approaches. For a
comparison, the form factor predictions of the other approaches at
$q^2=0$ are shown in Table. \ref{TT33}. The results of other methods have been rescaled
according to the form factor definition in Eq. (\ref{eq23}). It is useful to present the relations between our form factors ($A$, $V_i$) in Eq. ({\ref{eq23}}) to those used in \cite{CLF04,QM99,Aliev99,LCSR07}. The relations read
\begin{eqnarray*}
A&=&\frac{(m_B+m_{a_1})}{(m_B-m_{a_1})}A^{[2]}=- A^{[3]},~~~~~~
V_0=-\frac{(m_B+m_{a_1})}{2m_{a_1}}V_0^{[2,3]},\nonumber\\
V_1&=&V^{[2]}_1=-\frac{(m_B+m_{a_1})}{(m_B-m_{a_1})}V^{[3]}_1,~~~~~~
V_2=-\frac{(m_B+m_{a_1})}{(m_B-m_{a_1})}V^{[2]}_2= V_2^{[3]}.
\end{eqnarray*}
Also, the relation between our form factors to those used in \cite{LCSR07} and \cite{Aliev99} are obtained from the above equations by replacing
$A^{[3]}\to -A^{[4]},~ V^{[3]}_i\to -V^{[4]}_i$ and, $A^{[3]}\to \kappa A^{[5]}$, $V_i^{[3]}\to \kappa V_i^{[5]}$ respectively, where $\kappa=\frac{\sqrt{2}~m_{a_{1}} }{g_{a_{1}}f_{a_{1}}}$.
\begin{table}[th]
\caption{ Transition form factors of the $B\to a_1 \ell \nu$ at $q^2=0$
in various models. The results of other methods have been rescaled
according to the form factor definition in Eq. (\ref{eq23}).}
\label{TT33}
\begin{ruledtabular}
\begin{tabular}{ccccc}
Model&${A}(0)$&${V}_{0}(0)$&${V}_{1}(0)$&${V}_{2}(0)$\\
\hline
LFQM\cite{CLF04}& $0.67$&$0.34$ &  $0.37$ & $-0.29$   \\
CQM \cite{QM99} &$0.23$& $3.11$ &$1.32$ & $-0.55$   \\
LCSR\cite{LCSR07}&$0.48\pm0.09$&$0.77\pm0.13$&$0.60\pm0.11$&$-0.42\pm0.08$\\
SR \cite{Aliev99} &$0.55\pm0.08$&$0.49\pm0.11$&  $0.56\pm0.07$ & $-0.43\pm0.04$  \\
This Work&$0.51\pm0.11$&$0.46\pm0.10$&$0.52\pm0.11$&$-0.41\pm0.09$
\end{tabular}
\end{ruledtabular}
\end{table}

The errors in Table. \ref{TT33} are estimated by the variation of the Borel parameters
$M_1^2$ and $M_2^2$, the variation of the continuum thresholds
$s_0$ and $s_0^\prime$, the variation of $b$ quark mass
and leptonic decay constants $ f_{B}$ and $f_{a_1}$. The
main uncertainty comes from the thresholds and the decay
constants, which is about $\sim 25\% $ of the central value, while
the other uncertainties are small, constituting a few percent.

The dependence of the form factors, $A^{(1)}, V^{(1)}_{i}$, $T^{(1)}_{j}(q^2)$ and
$A^{(2)}, V^{(2)}_{i}$, $T^{(2)}_{j}$ on $q^2$ extracted from the fit functions,
Eqs. ({\ref{eq51}) and ({\ref{eq52}), are given in Figs. (\ref{F31})
and (\ref{F32}), respectively.
\begin{figure}
\includegraphics[width=6cm,height=5cm]{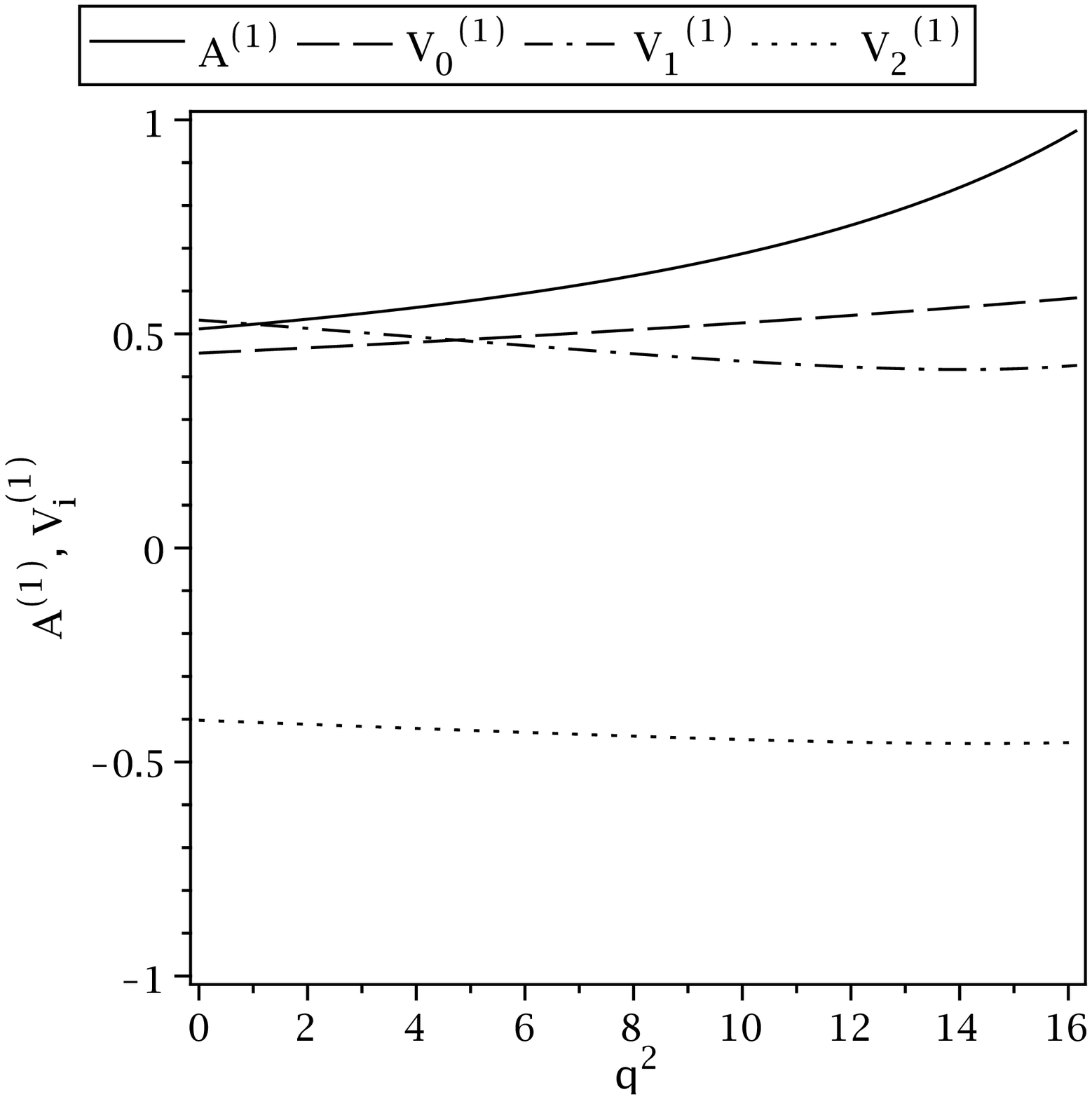}
\includegraphics[width=6cm,height=5cm]{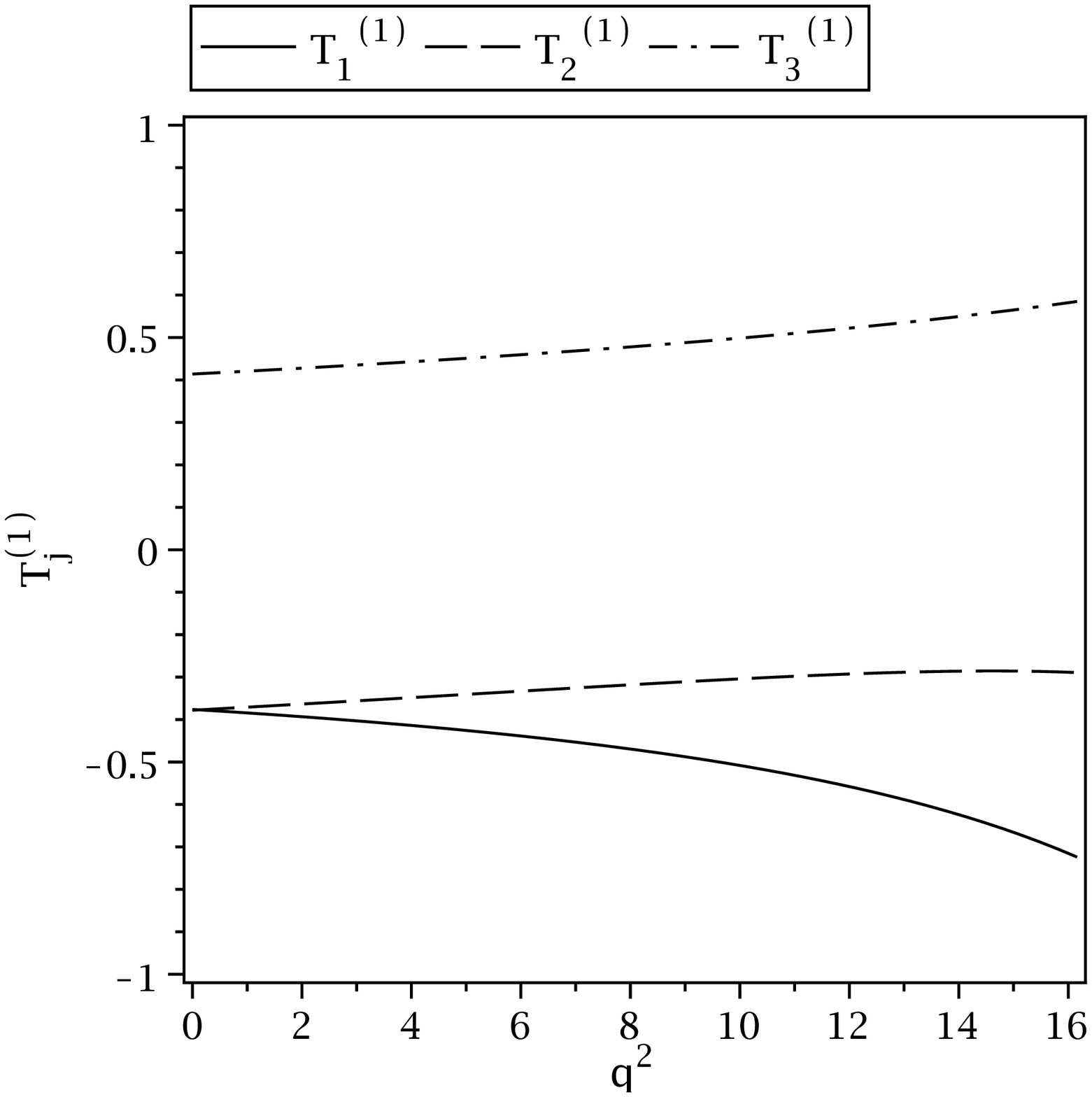}
\caption{The form factors $A^{(1)}, {V}^{(1)}_{i}$ and ${T}^{(1)}_{j}$ on $q^2$.} \label{F31}
\end{figure}
\begin{figure}
\includegraphics[width=6cm,height=5cm]{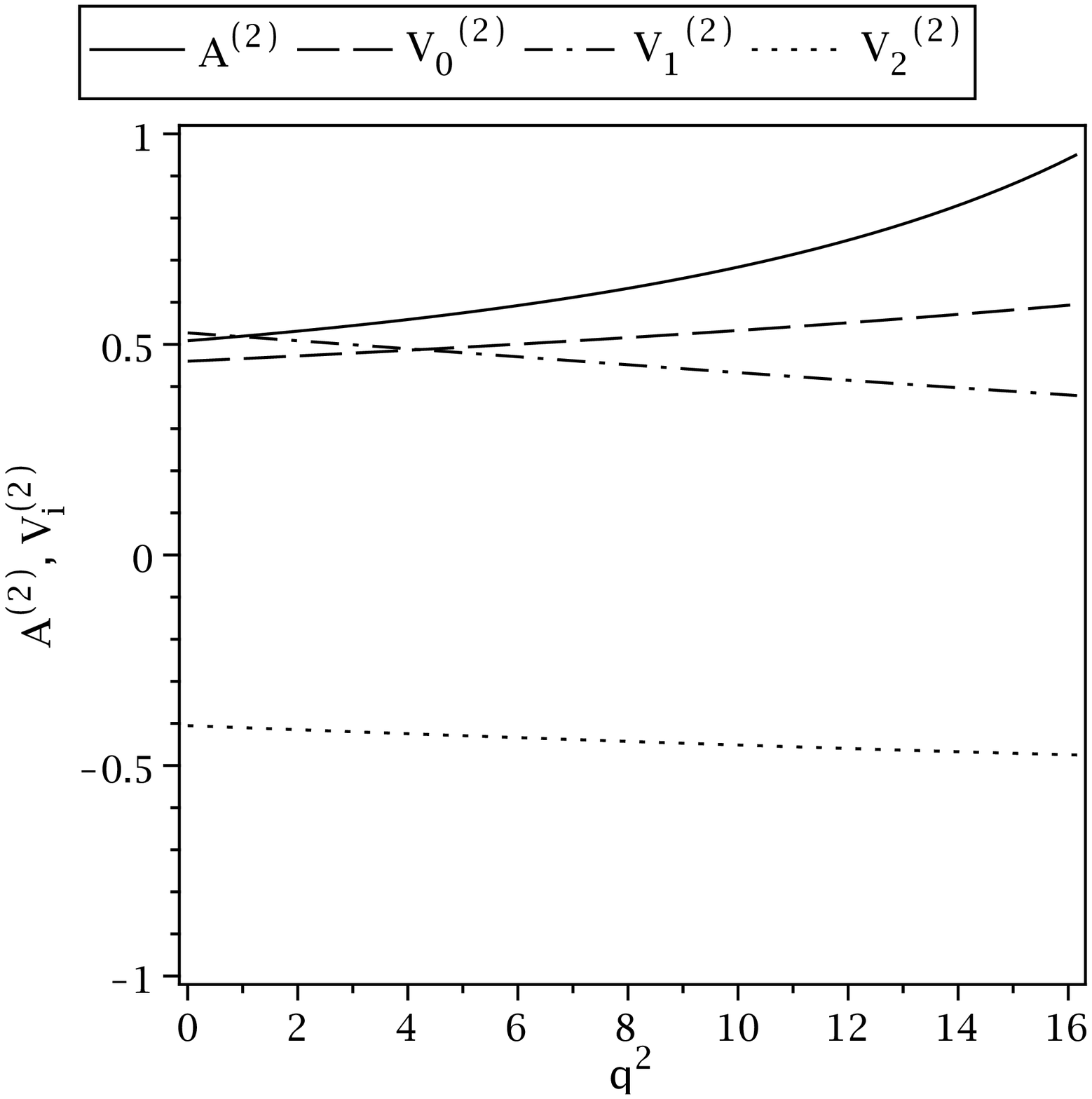}
\includegraphics[width=6cm,height=5cm]{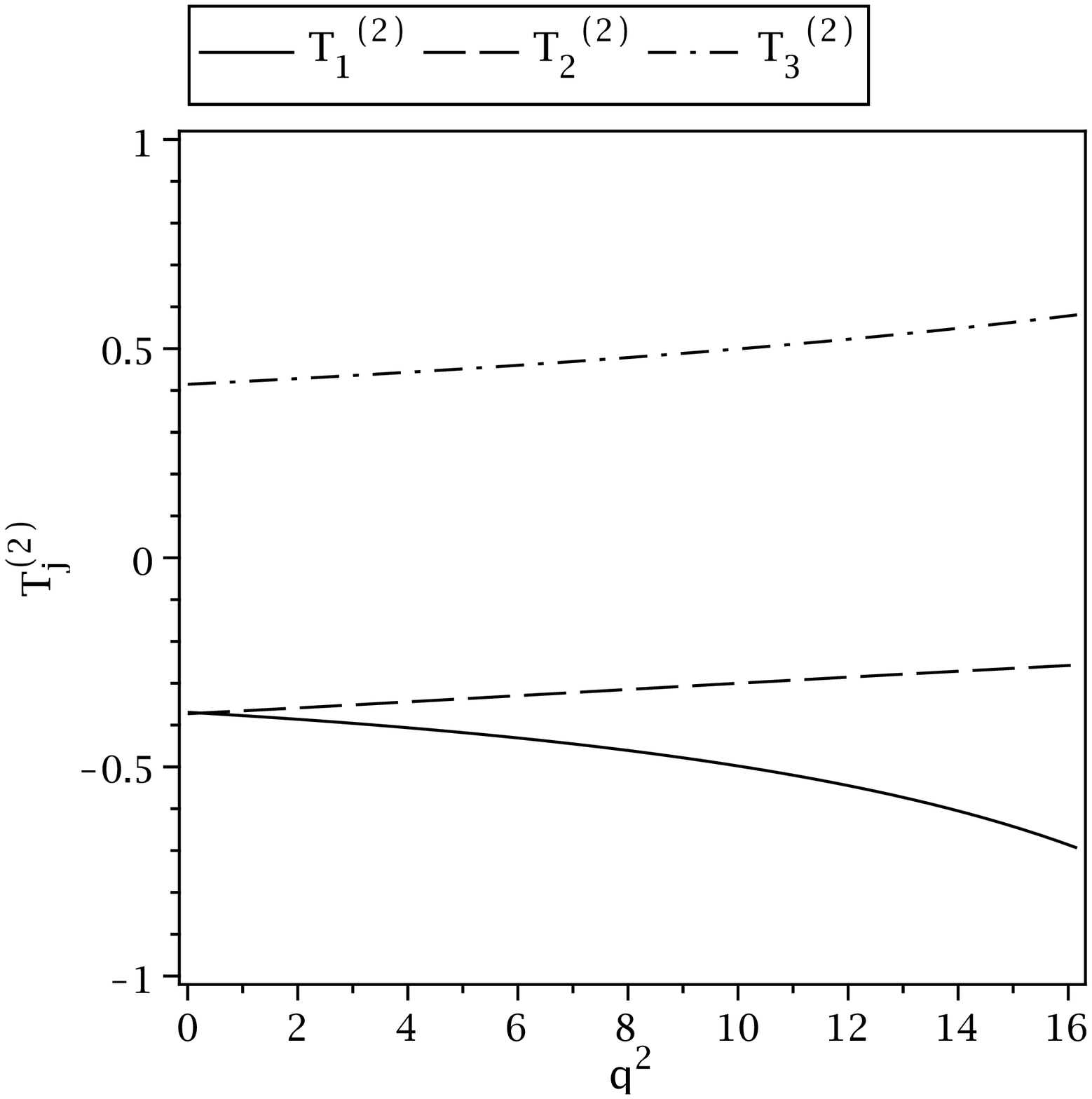}
\caption{The form factors $A^{(2)}, {V}^{(2)}_{i}$ and ${T}^{(2)}_{j}$ on $q^2$.} \label{F32}
\end{figure}

In the standard model, the rare semileptonic $B \to a_{1}
\ell^{+}\ell^{-}$ and $B \to \rho \ell^{+}\ell^{-}$ decays are
described  via  loop transitions, $b \rightarrow d~\ell^{+}\ell^{-}$ at
quark-level. Both mesons $a_1$ and $\rho$ have the same quark
content, but different masses and parities ,i.e., $\rho$ is a vector
$(1^-)$ and $a_1$ is a axial vector $(1^+)$.
We have calculated the form factor values of the $B\to \rho \ell^{+} \ell^{-}$ at $q^2=0$ in the SR model shown in Table. \ref{TTT33}.
Also, this table contains the results estimated for these form factors in the frame work of the LCSR.
The predicted values by us  and the LSCR model are very close to each other in many cases.
If $a_1$ behaves as the scalar partner of the $\rho$ meson, it
is expected that the $A(0)$ for the $B\to a_1$ decays is similar to the $V(0)$ for the $B\to \rho$ transitions, for example.
The values obtained for $A(0)$ via two the SR and LCSR models in Table. \ref{TT33}
are larger than those for $V(0)$ in Table. \ref{TTT33}.
It appears to us that the transition form factors of the $B \to a_{1}$  decays are quite different of those for $B \to \rho$.
\begin{table}[tbp]
\caption{The form factor values of the $B\to \rho \ell^+ \ell^-$ at $q^2=0$. }\label{TTT33}
\begin{ruledtabular}
\begin{tabular}{ccccccccc}
Mode & $V(0)$ & $A_0(0)$ &  $A_1(0)$ &  $A_2(0)$ & $T_1(0)$& $T_2(0)$ &  $T_3(0)$   \\
\hline
\mbox{This Work}&$0.30\pm0.09$ & $0.29\pm0.08$ &  $0.24\pm0.06$ & $0.20\pm0.07$ & $0.26\pm0.07$&  $0.26\pm0.07$ & $0.16\pm0.05$ \\
LCSR\cite{Ball}&$0.32$ & $0.30$ &  $0.24$ & $0.22$ & $0.27$&  $0.27$ & $0.18$
\end{tabular}
\end{ruledtabular}
\end{table}

Now, we would like to evaluate  the branching ratio values for the
$B\to a_1 \ell^+ \ell^-$ decays. The expressions of the differential decay width
$d\Gamma/dq^2$ for the $B \to a_1 \nu\bar{\nu}$ and $B \to a_1
\ell^+ \ell^-$ decays can be found in \cite{GL, Khosravi}. These
expressions contain the Wilson coefficients $C_{ 7}^{\rm eff}$,
$C_{9}^{\rm eff}$, $C_{10}$,  and also the CKM matrix elements $V_{tb}$
and $V_{td}$. Considering $C^{\rm eff}_{7}=-0.313$, $C_{10}=-4.669$,
$\mid V_{tb}V^{*}_{td}\mid=0.008$ \cite{FGIKL}, and the form factors
related to the fit functions, Eqs. ({\ref{eq51}) and ({\ref{eq52}),
and after numerical analysis,  the branching ratios for the $B \to
a_1 \ell^{+}\ell^{-}/\nu\bar{\nu}$  are obtained as presented in
Table \ref{T33}. In this table, we show only the values obtained  considering
the SD effects contributing to the Wilson
coefficient $C_9^{\rm eff}$ in Eq. (\ref{eq33}) for charged lepton case.
\begin{table}[th]
\caption{The branching ratios of the semileptonic $B\to a_1
\ell^+\ell^-$ decays, considering two groups of the form factors. $1$ and $2$ stand for the form factors, $F^{(1)}$ and  $F^{(2)}$, respectively.}\label{T33}
\begin{ruledtabular}
\begin{tabular}{ccccc}
\mbox{Mode}&\mbox{form factors}&\mbox{Value}\\
\hline
\mbox{Br}$(B \to a_1 \nu\bar{\nu})\times 10^{8}$&$^{1}_{2}$&$_{7.78\pm2.32}^{7.41\pm2.44}$\\
\mbox{Br}$(B \to a_1  e^{+} e^{-})\times 10^{8}$&$^{1}_{2}$&$_{2.90\pm0.95}^{2.75\pm0.58}$\\
\mbox{Br}$(B \to a_1  \mu^+ \mu^-)\times 10^{8}$&$^{1}_{2}$&$_{2.70\pm0.89}^{2.54\pm0.47}$\\
\mbox{Br}$(B \to a_1 \tau^+ \tau^-)\times 10^{9}$&$^{1}_{2}$&$_{0.33\pm0.10}^{0.37\pm0.09}$
\end{tabular}
\end{ruledtabular}
\end{table}

In this part, we would like to  present the branching ratio values including LD effects via $C_9^{\rm eff}$.
Due to in our calculations $q^2 < m^2_{\psi(4040)}$,
we introduce some cuts
around the narrow resonances of the $J/\psi$ and $\psi^{\prime }$,  and study the
following three regions for muon:
\begin{eqnarray}\label{eq34}
\mbox{I}: &&\ \ \ \ \ \ \ \ 2 m_{\mu} \;\leq\; \sqrt{q^2}
\;\leq\; M_{J/\psi }-0.20\,,
\nonumber\\
\mbox{II}: && M_{J/\psi}+0.04 \;\leq\; \sqrt{q^2} \;\leq\;
M_{\psi^{\prime}}-0.10\,,
\nonumber \\
\mbox{III}: &&\ \ M_{\psi^{\prime}}+0.02 \;\leq\; \sqrt{q^2}
\;\leq\; m_{B}-m_{a_1}\,,
\end{eqnarray}
and the following two for tau:
\begin{eqnarray}\label{eq35}
\mbox{I}: & \ \ \ \ \ \ \ 2 m_{\tau} \;\leq\; \sqrt{q^2}
\;\leq\; M_{\psi'} - 0.02\,,
\nonumber\\
\mbox{II}: & M_{\psi'} + 0.02\; \leq\; \sqrt{ q^2}\; \leq\;
m_{B}-m_{a_1}\,.
\end{eqnarray}
In Table \ref{T34}, we present the
branching ratios for muon and tau obtained using the regions shown
in Eqs. (\ref{eq34}-\ref{eq35}), respectively. In our calculations, two groups of the form factors are considered. Here, we
should also stress that the results obtained for the electron are
very close to the results of the muon and for this reason, we only
present the branching ratios for muon in our table.
\begin{table}[th]
\caption{The branching ratios of the semileptonic $B\to a_1
\ell^+\ell^-$ decays including  LD effects in three regions. $1$ and $2$ stand for the form factors, $F^{(1)}$ and  $F^{(2)}$, respectively.
}\label{T34}
\begin{ruledtabular}
\begin{tabular}{cccccccc}
\mbox{Mode}&\mbox{form factors}&I&II&III&I+II+III\\
\hline
\mbox{Br}$(B \to a_1\mu^+\mu^-)\times10^8$&$^{1}_{2}$&$_{2.30\pm0.76}^{2.07\pm0.68}$&$_{0.26\pm0.09}^{0.27\pm0.09}$&$_{0.07\pm0.03}^{0.08\pm0.03}$&$_{2.63\pm0.88}^{2.42\pm0.80}$\\
$\mbox{Br}(B \to
a_1\tau^+\tau^-)\times10^{9}$&$^{1}_{2}$&$^{\mbox{undefined}}_{\mbox{undefined}}$&$_{0.10\pm0.03}^{0.11\pm0.04}$&$_{0.13\pm0.04}^{0.15\pm0.05}$&$_{0.23\pm0.07}^{0.26\pm0.09}$
\end{tabular}
\end{ruledtabular}
\end{table}
Considering the form factors, $F^{(1)}$ and  $F^{(2)}$, the dependency of the differential branching ratios on
$q^2$ with and without  LD effects for charged lepton case is shown in Fig. (\ref{F33}). In this figure, the solid and dash-dotted lines show the results without and with the LD effects, respectively, using the form factors, $F^{(1)}$. Also the circles and stars are the same as those lines but considering $F^{(2)}$.
\begin{figure}
\includegraphics[width=6cm,height=5cm]{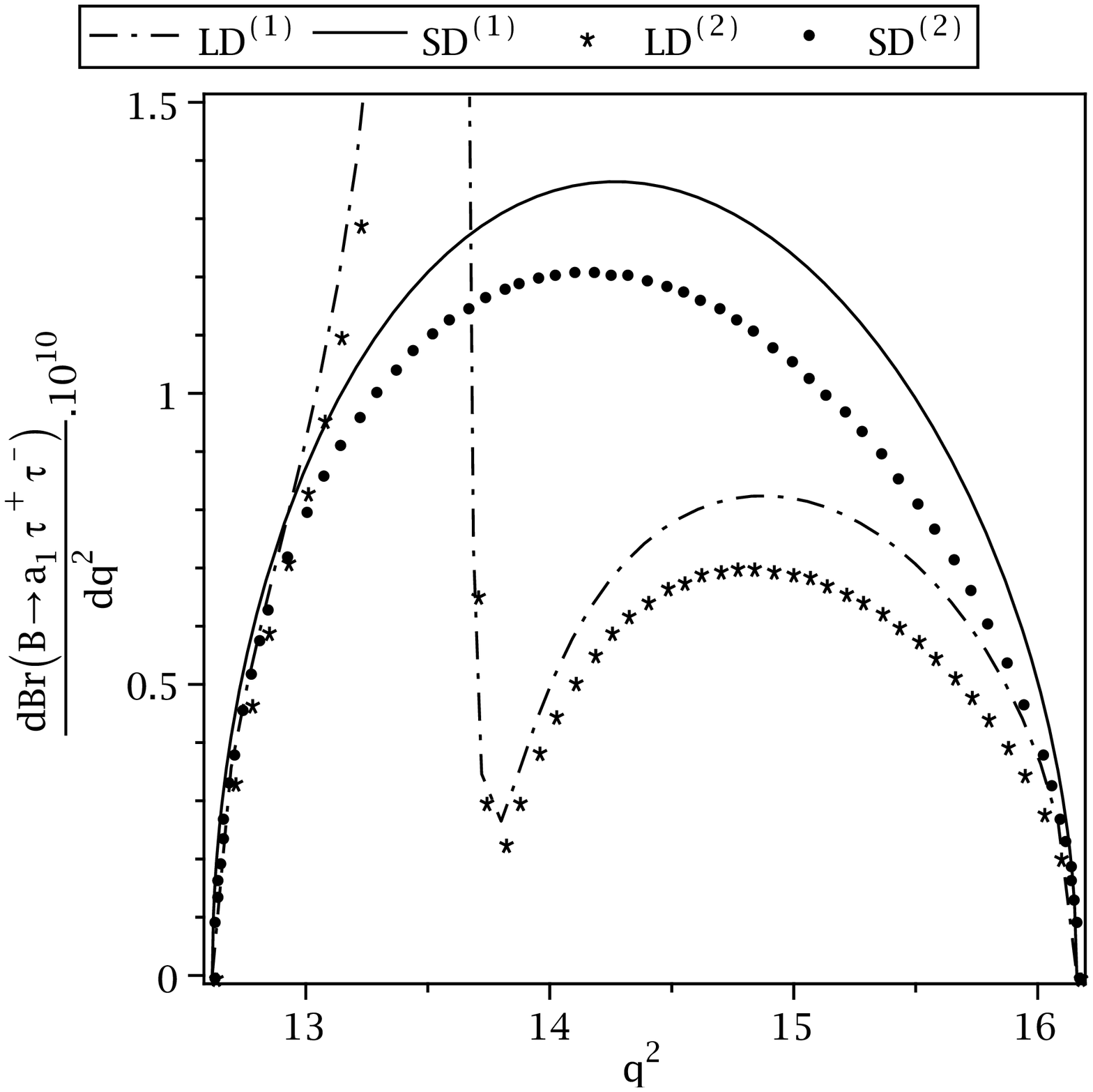}
\includegraphics[width=6cm,height=5cm]{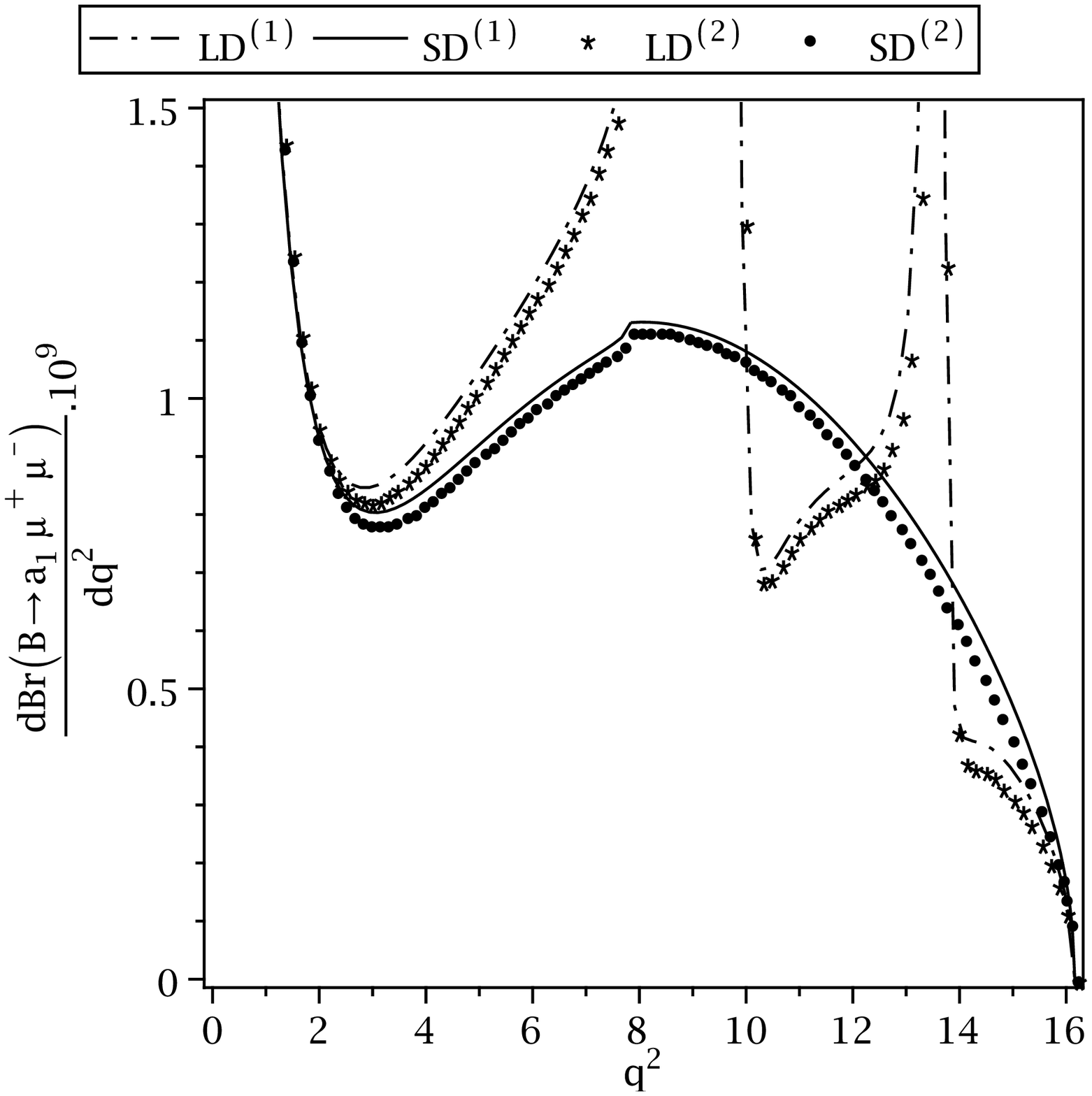}
\caption{The differential branching ratios of the semileptonic $B\to a_1$ decays on $q^2$ with and without  LD effects. }
\label{F33}
\end{figure}
In Ref. \cite{LYON}, the interference pattern of the charm-resonances $J/\psi(3370,4040,4160,4415)$  with the electroweak penguin operator $O_{9}$ in the branching fraction of $B^+ \to K^+ \mu^+ \mu^-$
has been investigated (in this case $q^2\simeq 22$~ GeV$^2$). For this purpose, the charm vacuum polarisation via a standard dispersion relation from BESII-data on $e^+e^- \to {\rm hadrons}$ is extracted. In the factorisation approximation the vacuum polarisation describes the interference fully non-perturbatively. The observed interference pattern by the LHCb collaboration is opposite in sign and significantly enhanced as compared to factorisation approximation. A change of the factorisation approximation result by a factor of $-2.5$, which correspond to a $350\%$-corrections, results in a reasonable agreement with the data.

Finally, we want to calculate  the longitudinal lepton polarization
asymmetry and the forward-backward asymmetry for the considered decays.
The expressions of the longitudinal lepton polarization
asymmetry and the forward-backward asymmetry, $P_L$ and $A_{FB}$,  are given in \cite{GL, Khosravi}:

The dependence of the longitudinal lepton polarization and the
forward-backward asymmetries  for the $B\to a_1 \ell^+\ell^-$ decays on
the transferred momentum square $q^2$ with and without LD effects
are plotted in Figs. (\ref{F34}) and (\ref{F35}), respectively.
\begin{figure}
\includegraphics[width=6cm,height=5cm]{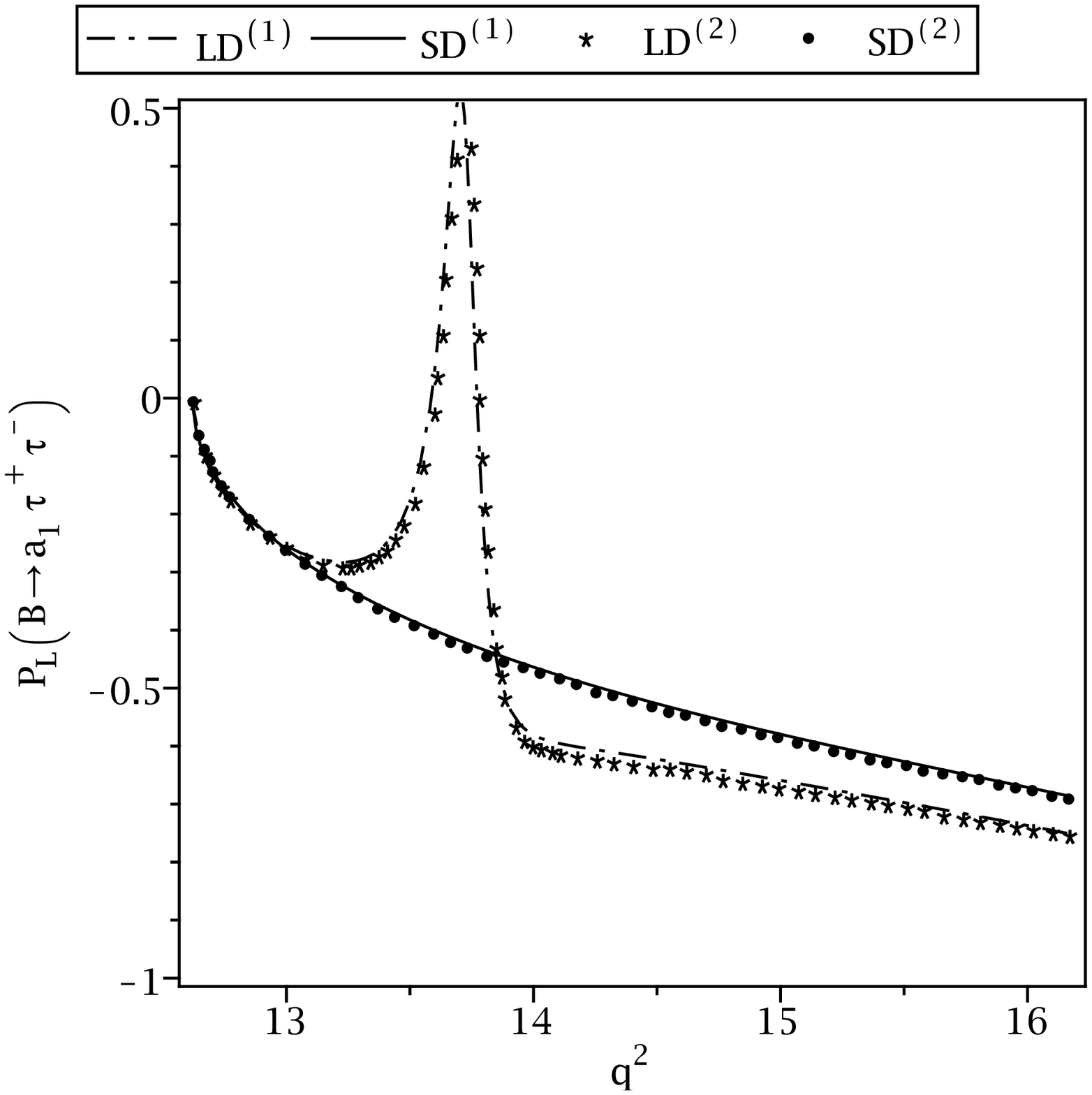}
\includegraphics[width=6cm,height=5cm]{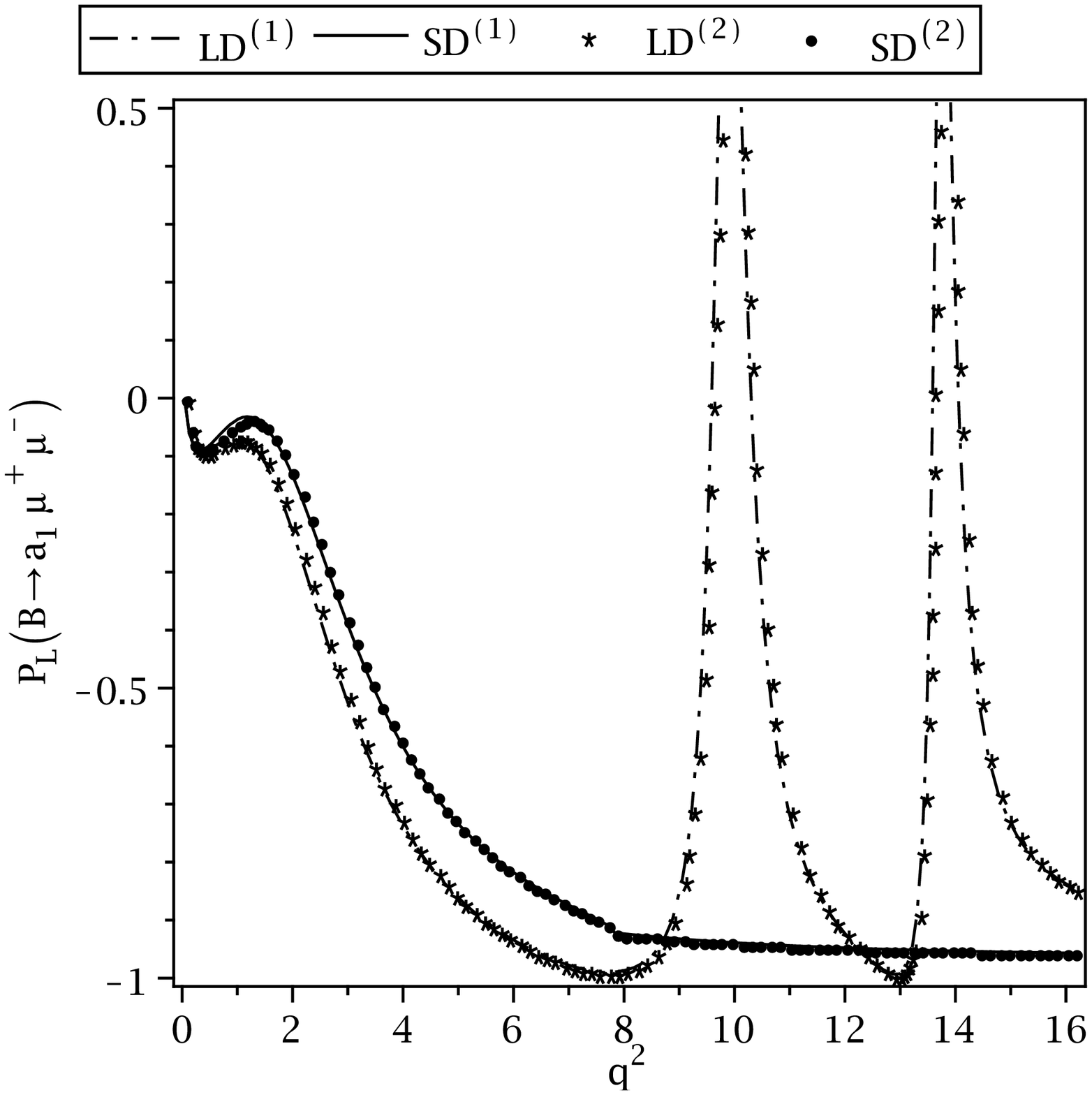}
\caption{The dependence of the longitudinal lepton polarization asymmetry on $q^2$
with and without the LD effects.}
\label{F34}
\end{figure}
\begin{figure}
\includegraphics[width=6cm,height=5cm]{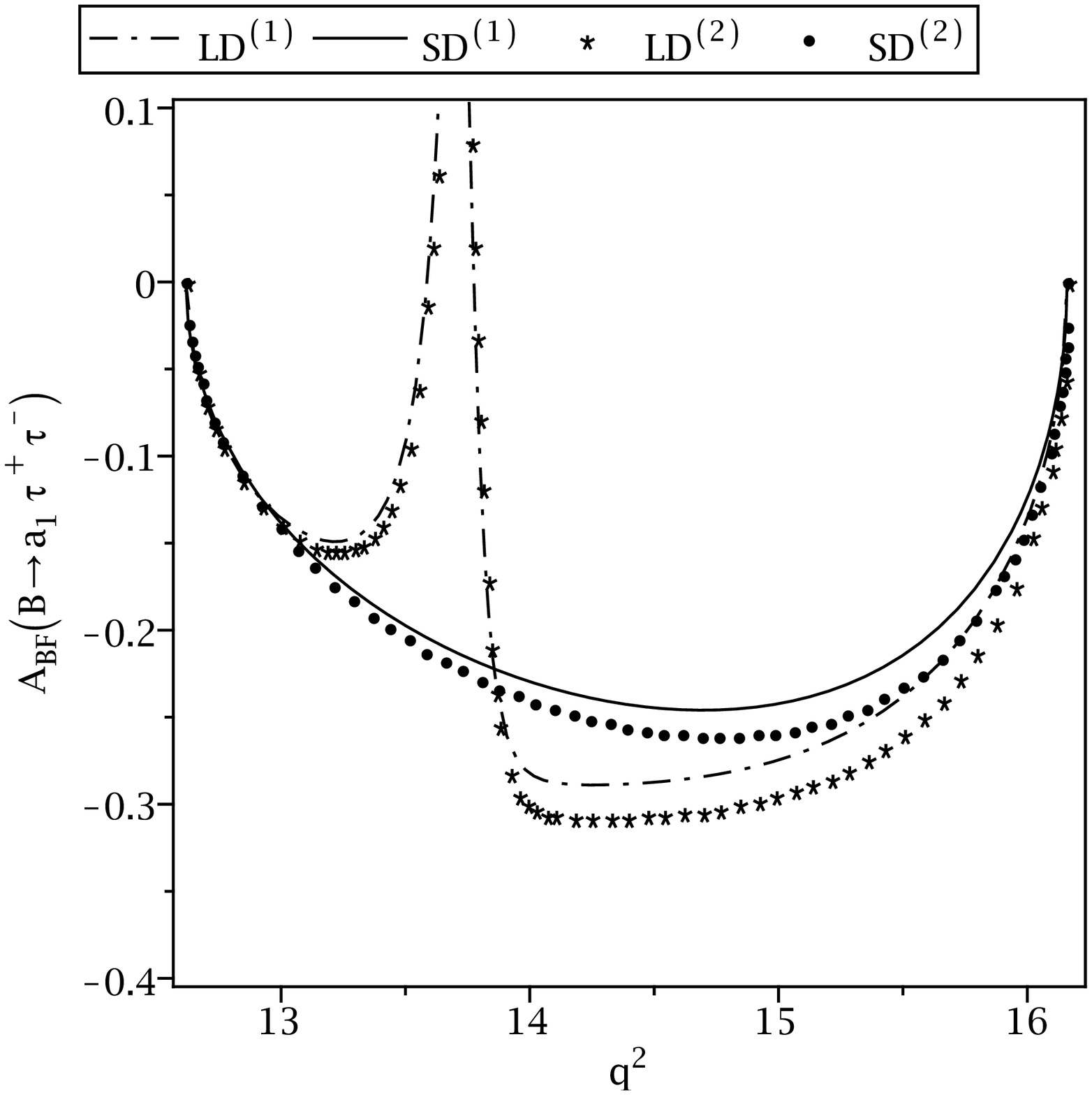}
\includegraphics[width=6cm,height=5cm]{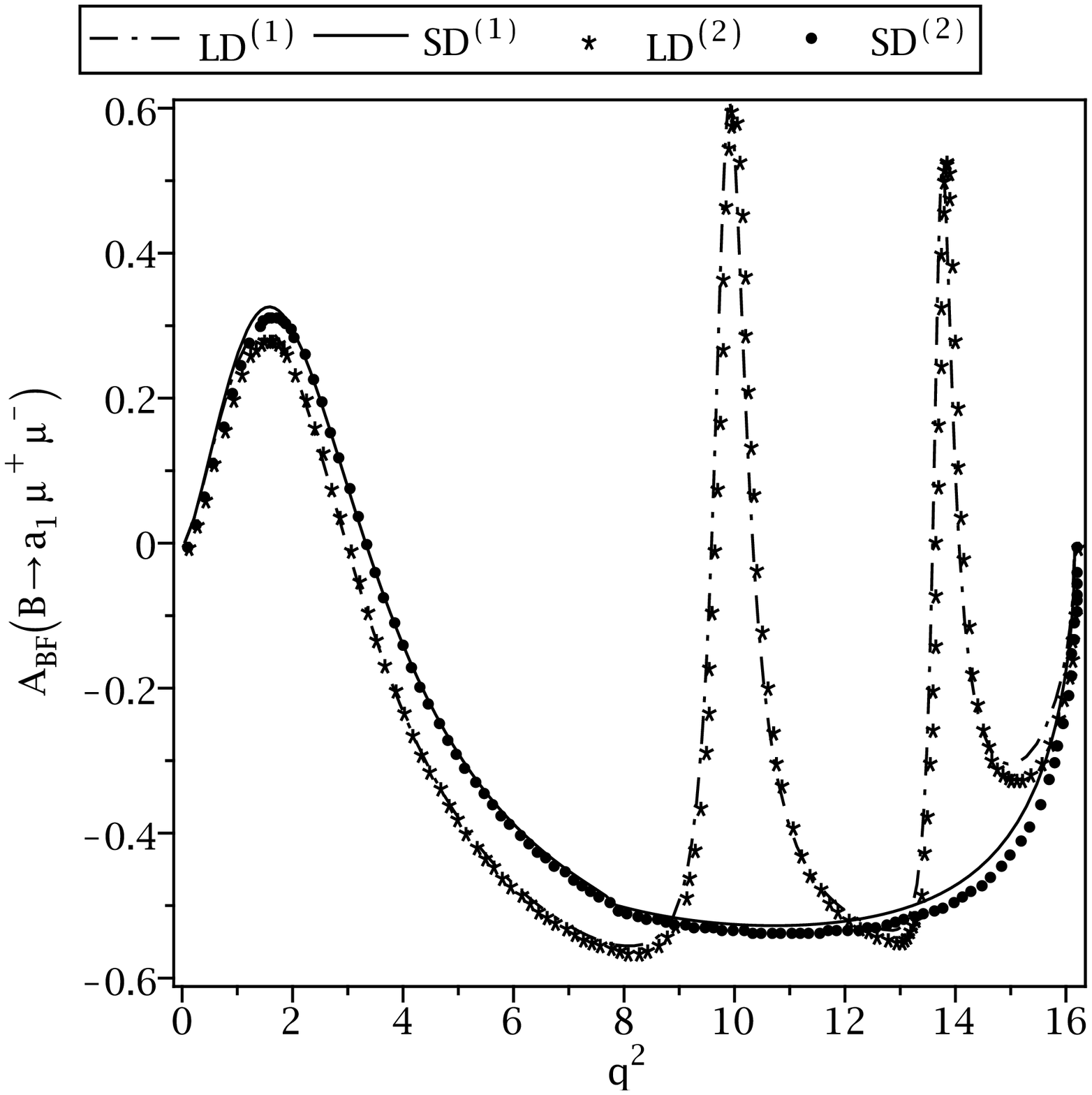}
\caption{The dependence of the forward-backward asymmetry on $q^2$
with and without the LD effects.}
\label{F35}
\end{figure}

The measurement of these quantities in the FCNC transitions are difficult. Among the large set of inclusive and exclusive FCNC
modes, a considerable attention has been put into $B\to K^* \mu^+\mu^-$such as:
measurement of the differential branching fraction and forward-backward asymmetry for
$B\to K^*  \ell^+\ell^-$ \cite{Wei},
measurements of the angular distributions in the decays $B\to K^*  \mu^+\mu^-$ \cite{Aaltonen},
differential branching fraction and angular
analysis of the decay $B\to K^*  \mu^+\mu^-$ \cite{Aaij}, Also angular distributions in the decay $B\to K^*  \ell^+\ell^-$ \cite{Aubert,GE}. In Ref. \cite{GE},  measurements of the BABAR are presented for the FCNC decayes, $B\to K^*  \ell^+\ell^-$ including branching
fractions, isospin asymmetries, direct CP violation, and lepton flavor universality for dilepton masses below and
above the $J/\psi$ resonance. Furthermore, BABAR results from an angular analysis in $B\to K^*  \ell^+\ell^-$ are reported in which
both the $K^*$ longitudinal polarization and the lepton forward-backward asymmetry are measured for dilepton
masses below and above the $J/\psi$ resonance.

In summary, the transition form factors  of
the semileptonic  $B \to a_1 \ell^+ \ell^-/\nu \bar{\nu}$ decays were
investigated in the 3PSR approach. Considering both the SD and LD effects contributing to the Wilson
coefficient $C_9^{\rm eff}$ for charged lepton case, we estimated the branching ratio values for these decays.
Also, for a better analysis, the dependence of the longitudinal
lepton polarization and forward-backward asymmetries of these decays on
$q^2$ were plotted.

\section*{Acknowledgments}
I would like to thank M. Haghighat for his useful discussions.
Partial support of the Isfahan University of Technology research
council is appreciated.

\end{document}